\documentclass[runningheads]{llncs}
\usepackage{graphicx}
\usepackage{hyperref}
\usepackage{tabularx}
\usepackage[export]{adjustbox}
\usepackage{hyperref}
\usepackage[hyphenbreaks]{breakurl}
\begin{document}

\title{AfricAIED 2024: 2nd Workshop on Artificial Intelligence in Education in Africa}

\titlerunning{AfricAIED 2024}


\author{George Boateng\inst{1,2} \and
Victor Kumbol \inst{1,3}}
\authorrunning{Boateng and Kumbol}

\institute{Kwame AI Inc.\\ 
\and
ETH Zurich, Switzerland\and
Charité -Universitätsmedizin Berlin, Germany}



\maketitle   










\begin{abstract}
Recent AI advancements offer transformative potential for global education, yet their application often overlooks Africa's unique educational landscape. AfricAIED 2024 will address this gap, spotlighting efforts to develop AI in Education (AIED) systems tailored to Africa's needs. Building on the success of the inaugural workshop, AfricAIED 2024 will feature an online AI Hackathon focused on democratizing preparation for Ghana's National Science \& Maths Quiz (NSMQ). Participants will create open-source AI tools leveraging resources from the Brilla AI project to level the academic playing field and enhance science and math education across Africa. The workshop will showcase top competitors' solutions, invite discussions on AIED opportunities and challenges in Africa, and highlight the latest advancements in AI education integration. AfricAIED 2024 aims to foster collaboration and innovation, amplifying African voices in the AIED community and driving positive change in African education through AI.


\end{abstract}



\section{Introduction}
Recent advances in AI systems such as BERT  \cite{devlin2019}and GPT-4\cite{achiam2023}, have demonstrated their capacity to transform education globally. Prominent figures in the EdTech industry, including Duolingo \cite{duolingomax}, Quizlet \cite{q-chat}, Chegg \cite{cheggmate}, and KhanAcademy \cite{khanmigo}, have actively incorporated these AI advancements into their platforms to enrich learning experiences. Nevertheless, the deployment and evaluation of these AI systems primarily focus on Western educational contexts, overlooking the distinct requirements and obstacles encountered by students in Africa. Notably, the introduction of GPT-4 in March 2023 featured various academic exams as benchmarks, none of which were from Africa \cite{achiam2023}. This omission underscores the tendency to marginalize Africa in the utilization of state-of-the-art AI innovations, disregarding the diverse educational environments and needs of African students. Consequently, the potential of AI to address educational disparities and hurdles prevalent in Africa remains largely unexplored, emphasizing the pressing need for greater inclusivity and customized solutions tailored to the specific challenges faced by students across the continent. AfricAIED — workshop on AI in Education in Africa — is the genesis of a movement to make pronounced the building and advancement of AIED systems that work well in the African context.

\section{Content and Themes}
Building upon a successful inaugural edition of this workshop last year — AfricAIED 2023  \cite{africaied2023} — this year's workshop — AfricAIED 2024 \footnote{\url{https://www.africaied.org}} — aims to crowdsource and highlight efforts to build and deploy AIED systems in Africa as well as discuss potential opportunities and challenges. This workshop will be centered around an online AI Hackathon that builds upon our open-source project, Brilla AI \cite{boateng2023nsmqai,boateng2024} which is building an AI Contestant to address our proposed grand challenge in education — \textbf{NSMQ AI Grand Challenge} — \textit{“Build an AI to compete in Ghana’s National Science \& Maths Quiz Ghana (NSMQ) competition and win — performing better than the best contestants in all rounds and stages of the competition”} \cite{boateng2023nsmq}. The NSMQ is an annual live science and mathematics competition for senior secondary school students in Ghana in which 3 teams of 2 students compete by answering questions across biology, chemistry, physics, and math in 5 rounds over 5 progressive stages until a winning team is crowned for that year \cite{NSMQ}. The NSMQ is an exciting live quiz competition with interesting technical challenges across speech-to-text, text-to-speech, question-answering, and human-computer interaction. An AI that conquers this grand challenge could have real-world impact on education such as enabling millions of students across Africa to have one-on-one learning support from this AI.

\section{AI Hackathon}
\subsection{Motivation}
There is a lot of inequity in preparations for the NSMQ where resources such as study materials, great teachers, etc, are available only to the big-name high schools that then almost always make it to the semis and finals,  and win. This problem is quite representative of the inequity in Ghana’s educational system. The goal of this challenge is to crowdsource AI-powered tools to democratize preparation for the NSMQ and give all schools a fairer chance at winning the NSMQ. More broadly, the tools could be extended to improve science and math learning across Africa.

\subsection{Challenge:} The AI hackathon challenge will be as follows:  “Build an open-source, AI-powered tool that enables students to prepare for the NSMQ using any of or a combination of the following open-source outputs from the Brilla AI project: (1) dataset consisting of quiz questions and open-source textbooks (2) code, (3) models for speech-to-text (STT) — transcribes speech with a Ghanaian accent, text-to-speech (TTS) — speaks with a Ghanaian accent (we have a model for the Quiz Mistress’ voice), and question-answering (QA). You are allowed to additionally use open-source code, models, and datasets. You are NOT allowed to use commercial models such as GPT-3.5.” Some potential ideas include an audio-based assessment system where a question is asked in the Quiz Mistress’ voice, accepts audio answers and says whether it’s correct or wrong along with an explanation, an audio-based question and answering system where you ask a question and get an answer and explanations, a multiplayer game of the quiz for assessment that tracks metrics on how quickly questions were answered, number of wrong answers, etc.

\subsection{Timeline and Participation:} It will run from May 15th to June 15th, 2024. Participants will be asked to make a submission by releasing their outputs with the Apache 2.0 license (so it is compatible with our open-source project) via GitHub and share (1) the link to the repo, (2) a 5-minute video explaining and demoing their system, and (3) a technical report (max 5 pages) describing the architecture of their technical system, and various design decisions made. The Brilla AI team will review the submissions and select the top 3 based on innovativeness and potential impact. They will receive cash prizes and be invited to present and demo their solution at AfricAIED 2024.

\section{Relevance and Importance to the AIED Community}
With the underrepresentation of AIED work from Africa, this workshop will highlight such work. It will increase participation from African scientists and engineers in the 2024 AIED conference. Also, the style of this workshop centered around a hackathon focused on AIED in Africa is likely to have a good number of contributions as there will be a lower barrier to participation compared to paper submissions as an example.

\section{Format and Activities}
AfricAIED 2024 will be run as a half-day workshop on Monday, 8th July 2024 in a hybrid format with the in-person component in Accra, Ghana. It will bring together educators, researchers, entrepreneurs, policymakers, and AI experts to discuss and collaborate on innovative ideas, best practices, and future developments in leveraging AI to enhance learning experiences and outcomes in Africa. The workshop will consist of technical presentations by the top 3 competitors on their winning approaches, invited talks, and panel discussions on opportunities for AIED, and challenges in developing and deploying AIED in Africa. The workshop will also feature presentations and demos from team leads of the Brilla AI team on the 2024 version of Brilla AI. The target audience is students from Africa, African researchers in the area of education, and researchers with an interest in AIED research in Africa. We expect to have a maximum of 50 participants in person and 50 participants online.

\section{Previous Editions of the Workshop: AfricAIED 2023}
We successfully ran the inaugural event of AfricAIED — 1st workshop on AI in Education in Africa — at Google Research Ghana with sponsorship from Google and ETH for Development. We had about 35 people in person and 40 online  \cite{africaied2023}. Present was also JoyNews, a National TV news broadcaster in Ghana. We had an incredible lineup of 16 speakers spanning educators, education researchers, education entrepreneurs, and AI experts who gave talks on the ways AI is already being used in education in Africa, a primer on AI, and Brilla AI, an AI being built to win Ghana's NSMQ. There was also a lively panel session that discussed the opportunities and challenges of leveraging AI to improve education in Africa. One of the key highlights was the demo of Brilla AI, the first version of our AI Contestant for the 2023 NSMQ that transcribed speech with a Ghanaian accent, generated an answer to a scientific riddle, and said that answer with a Ghanaian accent.

\section{Bios of Organizers}
\textbf{Dr. George Boateng} is a Computer Scientist, Engineer, Educator, and Social Entrepreneur recognized as one of the 2023 Forbes 30 Under 30 Europe, and 2021 MIT Technology Review’s 35 Innovators Under 35. He is a Postdoctoral Researcher (focus on Digital Biomarkers), Core Lead of Wearable AI for Rheumatoid Arthritis Management (WARAM), and Lecturer at ETH Zurich, Switzerland. His research spans ubiquitous computing (mobile and wearable), applied machine learning, mobile health, and education, and has resulted in over 30 peer-reviewed publications in international conferences and journals.  He is also the CEO and Cofounder of Kwame AI Inc., an AI start-up that empowers learners and knowledge workers (e.g., educators, researchers, lawyers) with their personal knowledge assistants to improve their outcomes and productivity significantly. He previously worked as a Visiting Researcher at the University of Cambridge, U.K., and as an Applied Scientist at Amazon (Alexa AI).  He has a BA in Computer Science and an MS in Computer Engineering from Dartmouth College, U.S., and a PhD in Applied Machine Learning from ETH Zurich, Switzerland.

\textbf{Victor Kumbol} is an early-stage neuroscientist, pharmacist and social entrepreneur currently pursuing a PhD in Medical Neurosciences at Charité -Universitätsmedizin Berlin. He is the COO and Cofounder of AI startup, Kwame AI and has been recognized on the 2023 Forbes 30 Under 30. As an innovator, Victor is passionate about leveraging technology to impact society and builds open labware for research.

\bibliographystyle{splncs04}
\bibliography{refs}

\begin{thebibliography}{10}
\providecommand{\url}[1]{\texttt{#1}}
\providecommand{\urlprefix}{URL }
\providecommand{\doi}[1]{https://doi.org/#1}

\bibitem{achiam2023}
Achiam, J., Adler, S., Agarwal, S., Ahmad, L., Akkaya, I., Aleman, F.L., Almeida, D., Altenschmidt, J., Altman, S., Anadkat, S., et~al.: Gpt-4 technical report. arXiv preprint arXiv:2303.08774  (2023)

\bibitem{africaied2023}
Africaied 2023 recap. \url{https://www.africaied.org/africaied-2023/recap}

\bibitem{boateng2023nsmq}
Boateng, G., Kumbol, V., Kaufmann, E.E.: Can an ai win ghana's national science and maths quiz? an ai grand challenge for education. arXiv preprint arXiv:2301.13089  (2023)

\bibitem{boateng2023nsmqai}
Boateng, G., Mensah, J.A., Yeboah, K.T., Edor, W., Mensah-Onumah, A.K., Ibrahim, N.D., Yeboah, N.S.: Towards an ai to win ghana’s national science and maths quiz. In: Deep Learning Indaba 2023 (2023)

\bibitem{boateng2024}
Boateng, G., Mensah, J.A., Yeboah, K.T., Edor, W., Mensah-Onumah, A.K., Ibrahim, N.D., Yeboah, N.S.: Brilla ai: Ai contestant for the national science and maths quiz. arXiv preprint arXiv:2403.01699  (2024)

\bibitem{cheggmate}
Chegg announces cheggmate, the new ai companion, built with gpt-4. \url{https://www.businesswire.com/news/home/20230417005324/en/Chegg-announces-CheggMate-the-new-AI-companion-built-with-GPT-4} (Mar 2023)

\bibitem{devlin2019}
Devlin, J., Chang, M.W., Lee, K., Toutanova, K.: {BERT}: Pre-training of deep bidirectional transformers for language understanding. In: Burstein, J., Doran, C., Solorio, T. (eds.) Proceedings of the 2019 Conference of the North {A}merican Chapter of the Association for Computational Linguistics: Human Language Technologies, Volume 1 (Long and Short Papers). pp. 4171--4186. Association for Computational Linguistics, Minneapolis, Minnesota (Jun 2019). \doi{10.18653/v1/N19-1423}, \url{https://aclanthology.org/N19-1423}

\bibitem{duolingomax}
Introducing duolingo max, a learning experience powered by gpt-4. \url{https://blog.duolingo.com/duolingo-max/} (Mar 2023)

\bibitem{khanmigo}
Harnessing gpt-4 so that all students benefit. a nonprofit approach for equal access. \url{https://blog.khanacademy.org/harnessing-ai-so-that-all-students-benefit-a-nonprofit-approach-for-equal-access/} (Mar 2023)

\bibitem{NSMQ}
National science and maths quiz. \url{https://nsmq.com.gh/}

\bibitem{q-chat}
Introducing q-chat, the world’s first ai tutor built with openai’s chatgpt. \url{https://quizlet.com/blog/meet-q-chat} (Mar 2023)

\end{thebibliography}


\end{document}